\documentclass[letterpaper, 10 pt, conference]{ieeeconf}

\IEEEoverridecommandlockouts
\usepackage{epsfig} 
\usepackage{amsmath} 
\usepackage{amssymb}  
\usepackage{bm}

\title{\LARGE \bf
Convergence in On-line Learning of Static and Dynamic Systems
}
\author{Torbj\"{o}rn Wigren, Ruoqi Zhang and Per Mattsson 
\thanks{This work was supported by the Swedish Research Council (VR) under
contract 2023-04546.}
\thanks{T. Wigren, R. Zhang and P. Mattsson are with the Department of Information Technology, Uppsala University, SE-75105 Uppsala, Sweden.
        {\tt\small \{torbjorn.wigren,ruoqi.zhang,per.mattsson\} @it.uu.se}.}%
}
\begin{document}

\maketitle
\thispagestyle{empty}
\pagestyle{empty}

\begin{abstract}
The paper derives analytical expressions for the asymptotic average updating direction of the adaptive moment generation (ADAM) algorithm when applied to recursive identification of nonlinear systems. It is proved that the standard hyper-parameter setting results in the same asymptotic average updating direction as a diagonally power normalized stochastic gradient algorithm. With the internal filtering turned off, the asymptotic average updating direction is instead equivalent to that of a sign-sign stochastic gradient algorithm. Global convergence to an invariant set follows, where a subset of parameters contain those that give a correct input-output description of the system. The paper also exploits a nonlinear dynamic model to embed structure in recurrent neural networks. A Monte-Carlo simulation study validates the results.
\end{abstract}

\section{INTRODUCTION}

The adaptive moment generation algorithm (ADAM) \cite{KingmaBa15} has become a workhorse in machine learning.
The majority of the many successful
applications use batch processing, sometimes with a large complexity and cost. In such cases recursiveness over time can  reduce the complexity to be linear in the amount of data. However, properties like convergence then needs to be analysed with
averaging, see \cite{Ljung77a} and \cite{KushnerClark78}. The paper therefore studies the asymptotic updating direction and convergence properties of a recursive variant of ADAM.

From a system identification point of view, the neural network, e.g. \cite{Prince23}, provides a parametrization that complements classical nonlinear model structures like piecewise linear static models  \cite{Aastrom85}, block oriented models \cite{StoicaSoderstrom82}, \cite{WestwickVerhaegen94}, the NARMAX class of models \cite{ChenBillings89}, as well as general state space  models \cite{SchonGutstavsson03}, \cite{Wigren05}. Some of the modern sequential Monte-Carlo (SMC) algorithms, like the bootstrap particle filter are also recursive \cite{Wigrenetal22}. The machine learning field has instead favoured general models like recurrent neural networks (RNNs) when applying so called supervised learning \cite{Prince23}. The advantage is generality, however the canonical model structures used in the system identification field minimize the set of parameters to avoid ambiguity and to maximize the accuracy, see \cite{LjungSoderstrom83}. Performance improvements can therefore be expected when structure is embedded into RNN models.

Convergence analysis of recursive identification algorithms was pioneered in \cite{Ljung77a}, \cite{KushnerClark78} and \cite{Ljung75}. These publications state how global stability of an averaged ordinary differential equation (ODE) associated with the algorithm is related to global convergence of the algorithm. The asymptotic paths of the recursive algorithm are also proved to converge to the solutions of the associated ODE, which enables the study of the average updating direction of ADAM performed in the paper. Recently, related results have been used to prove convergence of ADAM to a minimizer of the criterion function, see e.g. \cite{Reddietal18}, \cite{GadatGavra22}, \cite{Xiaoetal24}. No study focused on the average updating direction seems to have appeared.

The paper suggests a dynamic model that combines a discrete delay line chain and a static nonlinear function. When this nonlinear function equals a neural network, the result is an RNN with embedded structure which is believed to be a novel first contribution. The second contribution assumes a typical hyper-parameter setting of ADAM and then proves that the asymptotic average updating direction coincides with that of a diagonally power normalized stochastic gradient algorithm. Thirdly, with the internal filtering turned off, the asymptotic average updating direction is proved to correspond to that of a sign-sign stochastic gradient algorithm. The fourth contribution proceeds to prove that the algorithm converges globally to an invariant set, with a subset of parameters that give a correct input-output description of the system. The final Monte-Carlo simulation study verifies the theoretical results, and indicates correctness of the RNN.

The paper is organized as follows. Section II presents the structured RNN and the recursive version of ADAM. The convergence analysis appears in Sections III and IV. Experiments and conclusions follow in Sections V and VI.

\section{The Recursive Adaptive Moment Generation Algorithm}
\subsection{Model Structure}

To begin, a canonical nonlinear dynamic model is proposed to embed structure in RNNs. The model signals are the input signal vector $\mathbf{u}(t)$ consisting of $K$ vector signals, and the $n$-dimensional state vector $\hat{\mathbf{x}}(t, \bm{\theta})$, given by
\begin{equation}
\mathbf{u}(t)=\left( \mathbf{u}_1^T(t) \ \ ... \ \ \mathbf{u}_K^T(t)  \right)^T,
\label{eq:eq2}
\end{equation}
\begin{equation}
\mathbf{u}_k(t)=\left(u_k(t) \ \ ... \ \  u_k^{(n_k)}(t) \right)^T, \ k=1,...,K,
\label{eq:eq3}
\end{equation}
\begin{equation}
\hat{\mathbf{x}}(t, \bm{\theta}) = \left(  \hat{x}_1(t,\bm{\theta}) \ \ ... \ \ \hat{x}_n(t, \bm{\theta}) \right)^T.
 \label{eq:eq4}
\end{equation}
The superscript $^{(i)}$ denotes differentiation with respect to time $t$, $i$ times, to handle potential zero dynamics, and $\bm{\theta}$ denotes the unknown parameter vector with dimension $d$. The ODE underpinning the model then follows as
\begin{equation}
\left(
\begin{array}{c}
  \dot{\hat{x}}_1(t,\bm{\theta}) \\
  \vdots \\
  \dot{\hat{x}}_{n-1}(t,\bm{\theta}) \\
  \dot{\hat{x}}_n(t,\bm{\theta})
\end{array}
\right)
=
\left(
\begin{array}{c}
  \hat{x}_2(t,\bm{\theta}) \\
  \vdots \\
  \hat{x}_{n}(t,\bm{\theta}) \\
  f\left(\hat{\mathbf{x}}(t,\bm{\theta}),\mathbf{u}(t), \bm{\theta}   \right)
\end{array}
\right).
\label{eq:eq1}
\end{equation}
Here  $f\left(\hat{\mathbf{x}}(t,\bm{\theta}),\mathbf{u}(t), \bm{\theta}   \right)$  parameterizes the $n {\rm :th}$
component of the ODE. When selected as a neural network $f^{nn}\left(\hat{\mathbf{x}}(t,\bm{\theta}),\mathbf{u}(t), \bm{\theta}   \right)$ an RNN results. Using the Euler method to discretize (\ref{eq:eq1}) with sampling period $T_s$ gives
\begin{eqnarray*}
\hat{\mathbf{x}}(t+T_s,\bm{\theta})=
\left(
\begin{array}{c}
  \hat{x}_1(t+T_s,\bm{\theta}) \\
  \vdots \\
  \hat{x}_{n-1}(t+T_s,\bm{\theta}) \\
  \hat{x}_n(t+T_s,\bm{\theta})
\end{array}
\right)
\end{eqnarray*}
\begin{equation}
=
\left(
\begin{array}{c}
  \hat{x}_1(t,\bm{\theta}) \\
  \vdots \\
  \hat{x}_{n-1}(t,\bm{\theta}) \\
  \hat{x}_n(t,\bm{\theta})
\end{array}
\right)
+
T_s
\left(
\begin{array}{c}
  \hat{x}_2(t,\bm{\theta}) \\
  \vdots \\
  \hat{x}_{n}(t,\bm{\theta}) \\
  f\left(\hat{\mathbf{x}}(t,\bm{\theta}),\mathbf{u}(t), \bm{\theta}   \right)
\end{array}
\right).
\label{eq:eq0}
\end{equation}
The $p$-dimensional output measurement model is
\begin{equation}
\hat{\mathbf{y}}(t, \bm{\theta})=\mathbf{C}\hat{\mathbf{x}}(t, \bm{\theta}),
\label{eq:eq5}
\end{equation}
where $\mathbf{C}$ is the  measurement matrix. In case $n=0$, a nonlinear static model results. The parameterization of (\ref{eq:eq0}) is given by the details of $f\left(\hat{\mathbf{x}}(t,\bm{\theta}),\mathbf{u}(t), \bm{\theta}   \right)$, e.g. by the static neural network $f^{nn}\left(\hat{\mathbf{x}}(t,\bm{\theta}),\mathbf{u}(t), \bm{\theta}   \right)$ and its hyper-parameters.

The gradient of the model (\ref{eq:eq5}) is given by
\begin{equation}
\bm{\psi}^{\top}(t,\bm{\theta})=\frac{\partial \hat{\mathbf{y}}(t, \bm{\theta})}{\partial \bm{\theta}}=\mathbf{C} \frac{\partial \hat{\mathbf{x}}(t, \bm{\theta})}{\partial \bm{\theta}} =\mathbf{C}\bm{\Psi}(t,\bm{\theta}).
\label{eq:eq6}
\end{equation}
To obtain the matrix $\bm{\Psi}(t,\bm{\theta})$, the components of the difference equation (\ref{eq:eq0}) are differentiated with respect to $\bm{\theta}$ to give 
\begin{eqnarray*}
\left(
\begin{array}{c}
  \frac{\partial \hat{x}_1(t+T_s,\bm{\theta}) }{\partial \bm{\theta}} \\
  \vdots \\
  \frac{\partial \hat{x}_{n-1}(t+T_s,\bm{\theta})}{\partial \bm{\theta}} \\
  \frac{\partial \hat{x}_n(t+T_s,\bm{\theta})} {\partial \bm{\theta}}
\end{array}
\right)=\bm{\Psi}(t+T_s,\bm{\theta})
\end{eqnarray*}
\begin{equation}
=\bm{\Psi}(t,\bm{\theta})+T_s \left(
\begin{array}{c}
  \frac{\partial \hat{x}_2(t,\bm{\theta})}{\partial \bm{\theta}} \\
  \vdots \\
  \frac{\partial \hat{x}_{n}(t,\bm{\theta})}{\partial \bm{\theta} } \\
  \frac{\partial}{\partial \bm{\theta} }f\left(\hat{\mathbf{x}}(t,\bm{\theta}),\mathbf{u}(t), \bm{\theta}   \right)
\end{array}
\right).
\label{eq:eq6a}
\end{equation}
$\frac{\partial}{\partial \bm{\theta}}f\left(\hat{\mathbf{x}}(t,\bm{\theta}),\mathbf{u}(t), \bm{\theta}   \right)$ can be computed by auto-differentiation at run-time, cf. (\ref{eq:eq6a}) and ${\rm diff}_{\bm{\theta}}(\cdot)$ of (\ref{eq:eq10}).

\subsection{Algorithm}
When  ADAM  is applied to (\ref{eq:eq0})-(\ref{eq:eq6a}), there is a significant difference as compared to other recursive identification algorithms, since ADAM does not process the gradient of the model separately. Instead the full gradient of the criterion
\begin{eqnarray*}
V_A(\bm{\theta})=\frac{1}{2} \lim_{t\rightarrow \infty}E[\bm{\varepsilon}^{\top}(t,\bm{\theta})\bm{\varepsilon}(t,\bm{\theta})]
\end{eqnarray*}
\begin{equation}
=
\frac{1}{2} \lim_{t\rightarrow \infty}E[\left( \mathbf{y}(t)-\hat{\mathbf{y}}(t,\bm{\theta}) \right)^{\top}\left( \mathbf{y}(t)-\hat{\mathbf{y}}(t,\bm{\theta}) \right)]
\label{eq:eq7}
\end{equation}
is processed, where $\mathbf{y}(t)$ is the measurement.
The gradient that is approximated by ADAM is hence
\begin{equation}
\mathbf{g}(t, \bm{\theta})=\left( \frac{\partial V_A(\bm{\theta})}{\partial \bm{\theta}} \right)^{\top}=- \lim_{t \rightarrow \infty}E[\bm{\psi}(t,\bm{\theta})\bm{\varepsilon(t, \bm{\theta})}].
\label{eq:eq9}
\end{equation}
Hence, when ADAM estimates approximate second order properties, this is done for $\bm{\psi}(t,\bm{\theta})\bm{\varepsilon(t, \bm{\theta})}$ rather than for $\bm{\psi}(t,\bm{\theta})$ that would be the case for a Gauss-Newton based recursive identification algorithm, cf. \cite{LjungSoderstrom83}. This has significant consequences that will be discussed in Section IV.

After reordering the equations of Algorithm 1 of \cite{KingmaBa15} to coincide with \cite{Wigren23} to facilitate the convergence analysis, and after replacing model signals with running estimates, the recursive algorithm becomes
\begin{equation}
\begin{array}{rcl}
  \bm{\varepsilon}(t) & = & \mathbf{y}(t)-\hat{\mathbf{y}}(t) \\
  \mathbf{m}(t) & = & \beta_1\mathbf{m}(t-T_s) \\
  &  & +(1-\beta_1)(-\bm{\psi}(t)\bm{\varepsilon}(t))
  \\
  \hat{\mathbf{m}}(t) & = & \frac{\mathbf{m}(t)}{1-\beta_1^{{\rm round}(t/T_s)}} \\
  \mathbf{v}(t) & = & \beta_2\mathbf{v}(t-T_s) +(1-\beta_2)\\
  & & \times ({\rm \bf{vec}}({\rm \bf{diag}}(\bm{\psi}(t)\bm{\varepsilon}(t)\bm{\varepsilon}^{\top}(t)\bm{\psi}^{\top}(t))))
  \\
  \hat{\mathbf{v}}(t) & = & \frac{\mathbf{v}(t)}{1-\beta_2^{{\rm round}(t/T_s)}} \\
  \hat{\bm{\theta}}(t) & = & \hat{\bm{\theta}}(t-T_s)-\alpha(t) \left( \hat{\mathbf{v}}^{ \cdot \frac{1}{2}}(t)+\epsilon \right)^{ \cdot -1} \cdot \hat{\mathbf{m}}(t) \\
  \hat{\mathbf{x}}(t+T_s)
  & = & \left(
\begin{array}{c}
  \hat{x}_1(t) \\
  \vdots \\
  \hat{x}_{n-1}(t) \\
  \hat{x}_n(t)
\end{array}
\right)
\\
& & +
T_s
\left(
\begin{array}{c}
  \hat{x}_2(t) \\
  \vdots \\
  \hat{x}_{n}(t) \\
  f\left(\hat{\mathbf{x}}(t),\mathbf{u}(t), \hat{\bm{\theta}}(t) \right)
\end{array}
\right) \\
  \hat{\mathbf{y}}(t+T_s) & = & \mathbf{C}\hat{\mathbf{x}}(t+T_s) \\
\bm{\Psi}(t+T_s)  & = & \bm{\Psi}(t) \\
& & +T_s \left(
\begin{array}{c}
  \frac{\partial \hat{x}_2(t,\bm{\theta})}{\partial \bm{\theta}} \\
  \vdots \\
  \frac{\partial \hat{x}_{n}(t,\bm{\theta})}{\partial \bm{\theta} } \\
  {\rm diff}_{\bm{\theta}} \left(f\left( \hat{\mathbf{x}}(t),\mathbf{u}(t), \hat{\bm{\theta}}(t) \right) \right)
\end{array}
\right) \\
\bm{\psi}(t+T_s)  & = & \left(\mathbf{C}\bm{\Psi}(t+T_s)\right)^{\top}.
\end{array}
\label{eq:eq10}
\end{equation}
To explain (\ref{eq:eq10}) and its notation, it is first noted that the element-wise multiplication $\mathbf{g}(t)\odot \mathbf{g}(t)$ of \cite{KingmaBa15} may be re-written as a vectorizing operation on the matrix $ \rm{\bf{diag}} (\bm{\psi}(t)\bm{\varepsilon}(t)\bm{\varepsilon}^{\top}(t))\bm{\psi}^{\top}(t)$, where ${\rm \bf{diag}}(\cdot)$ extracts the diagonal matrix from a matrix and ${\rm \bf{vec}(\cdot)}$ creates a vector of the diagonal elements. The additional element-wise operations of ADAM use a notation with a dot $(\cdot)$ before the mathematical operation. When the operation is implicit like for multiplication, a dot means element-wise operation.

The quantities of the algorithm include $\mathbf{m}(t)$ which denotes the first (order) moment and ${\hat{\mathbf{m}}}(t)$ which denotes the bias corrected first moment, while $\mathbf{v}(t)$ and ${\hat{\mathbf{v}}}(t)$ denote the second (order) moment and the bias compensated counterpart, used to approximate a Newton search. $\beta_1$ and $\beta_2$ are filtering hyper-parameters with standard values $0.9$ and $0.999$, and $\alpha(t)\propto t^{-1}$ is the gain sequence that needs to replace the constant step size $\alpha$ in the convergence analysis.

\section{Averaging Analysis}
\subsection{The Associated ODE and Convergence Relations}
The analysis follows a similar path as \cite{Wigren23}. First regularity conditions are defined, so that the averaging results of \cite{Ljung77a} and \cite{Ljung75} can be re-used. The average updating direction can then be analysed, interpreted and compared to classical gradient descent algorithms, for different hyper-parameters. Global convergence of (\ref{eq:eq10}) finally follows from the Lyapunov stability of the associated ODE of (\ref{eq:eq10}). The argument $(t,\theta)$ indicates  a {\em fixed} $\theta$ when expectations are computed.

\subsection{Conditions}
The averaging analysis requires  regularity conditions, defined below. The conditions M1 - M4 define a model set for which exponential stability and ergodicity holds. This may seem restrictive, however projection algorithms need to enforce asymptotic stability of  simulated models and gradients in recursive identification of linear systems as well \cite{LjungSoderstrom83}. M2 restricts the scope to continuously differentiable activation functions in case a neural network is used in (\ref{eq:eq0}) and (\ref{eq:eq10}). The conditions A1, S1 and S2 imply that also the data generating system is exponentially stable. The condition G1 ensures an appropriate decay rate of the gain sequence.

A2 lists the quantities of the average updating direction. Referring to \cite{LjungSoderstrom83}, average updating directions for $\hat{\bm{\theta}}(t)$ and $\hat{\mathbf{v}}(t)$ are needed, both with gain sequences $\sim 1/t$ to fit the general algorithm of \cite{Ljung75}, cf.  equation (A1) of \cite{Wigren23}. To achieve this, the analysis for the first set of standard hyper-parameters need to be restricted by $\beta_2 \rightarrow 1$. When the bandwidth of the unity gain autoregressive filtering then tends to $0$, $\mathbf{v}(t)$ and consequently $\hat{\mathbf{v}}(t)$ approaches the expected value of the input to the autoregressive filter, and
\begin{eqnarray*}
\lim_{\beta_2 \rightarrow 1} \lim_{t \rightarrow \infty} \hat{\mathbf{v}}(t,\bm{\theta})=\lim_{\beta_2 \rightarrow 1} \lim_{t \rightarrow \infty} \mathbf{v}(t,\bm{\theta})
\end{eqnarray*}
\begin{eqnarray*}
=\lim_{t \rightarrow \infty} E \left[ {\rm \bf{vec}}({\rm \bf{diag}}(\bm{\psi}(t,\bm{\theta})\bm{\varepsilon}(t,\bm{\theta})\bm{\varepsilon}^{\top}(t,\bm{\theta})
\bm{\psi}^{\top}(t,\bm{\theta}))) \right]
\end{eqnarray*}
\begin{equation}
\label{eq:eq200}
=\lim_{t \rightarrow \infty} \frac{T_s}{t}\sum_{k=1}^{\frac{t}{T_s}} {\rm \bf{vec}}({\rm \bf{diag}}(\bm{\psi}(´kT_s)\bm{\varepsilon}(kT_s)\bm{\varepsilon}^{\top}(kT_s)
\bm{\psi}^{\top}(kT_s))).
\end{equation}
The first equality follows since $\left(1-\beta_2^{{\rm round}(t/T_s)}  \right)^{-1}$ $\rightarrow 1$ as $t \rightarrow \infty$. The equality with the sample average follows from ergodicity. The equation underpinning the sample average of  (\ref{eq:eq200}) can then be written recursively as
\begin{eqnarray*}
\mathbf{v}(t)=\mathbf{v}(t-T_s)
\end{eqnarray*}
\begin{equation}
\label{eq:eq201}
+\frac{T_s}{t}\left(
{\rm \bf{vec}}({\rm \bf{diag}}(\bm{\psi}(t)\bm{\varepsilon}(t)\bm{\varepsilon}^{\top}(t)
\bm{\psi}^{\top}(t)))-\mathbf{v}(t-T_s)
\right).
\end{equation}
The updating structure of (\ref{eq:eq10}) and (\ref{eq:eq201}) then appear in A2:
\begin{itemize}
\item[M1:] The system and model are single output, i.e. $p=1$.
\item[M2:] The model set $\mathcal{D}_\mathcal{M}$ is a compact subset of $\mathcal{R}^{d+d}$, such that $\left( \bm{\theta}^{\top} \ \ \mathbf{v}^{\top}  \right)^{\top} \in \mathcal{D}_\mathcal{M}$ implies continuously differentiable, exponentially stable and bounded state dynamics, state gradient dynamics, and derivatives.
\item[M3:] $\left( \bm{\theta}^{\top} \ \ \mathbf{v}^{\top} \right)^{\top} \in \mathcal{D}_\mathcal{M}$ implies that $\mathbf{v}(t)>\delta_\mathbf{v}\mathbf{1}$, $\delta_\mathbf{v}>0$.
\item[M4:] $\mathbf{u}(t)=\left( u_1(t) \ \ ... \ \ u_K(t) \right)^{\top}$, without time derivatives, is generated from i.i.d bounded random vectors $\{\bar{\mathbf{u}}(t) \}$, by asymptotically stable linear filtering.
\item[G1:] $\lim_{t \rightarrow \infty} t \alpha(t)= \bar{\alpha}$, $0<\bar{\alpha}<\infty$.
\item[A1:] The data  $\{ \mathbf{z}(t) \} = \{ \left( y(t) \ \ \mathbf{u}^{\top}(t) \right)^{\top} \}$ is strictly stationary, ergodic and $\| \mathbf{z}(t) \| \leq C<\infty$, w.p.1, $\forall t$.
\item[A2:] The following limits exist for $\left( \bm{\theta}^{\top} \ \ \mathbf{v}^{\top}  \right)^{\top} \in \mathcal{D}_\mathcal{M}$ when $\beta_2 \rightarrow 1$:
    \begin{eqnarray*}
    \mathbf{f}(\bm{\theta}, \mathbf{v})= \lim_{t \rightarrow \infty} E\left[ \left( \mathbf{v}^{\cdot \frac{1}{2}} +\epsilon\mathbf{I} \right)^{\cdot -1} \cdot \hat{\mathbf{m}}(t,\bm{\theta}) \right],
    \end{eqnarray*}
    \begin{eqnarray*}
    \mathbf{G}(\bm{\theta})=\lim_{t \rightarrow \infty} E \left[
    {\rm \bf{vec}}({\rm \bf{diag}}(\bm{\varepsilon}^2(t,\bm{\theta}) \bm{\psi}(t,\bm{\theta})
   \bm{\psi}^{\top}(t,\bm{\theta}))) \right].
    \end{eqnarray*}
\item[S1:]For each $t,s, t\geq s$, there exists a random vector $\mathbf{z}_s^0(t)$ that belongs to the $\sigma$-algebra generated by $\mathbf{z}^t$ but is independent of $\mathbf{z}^s$ (for $s=t$ take $\mathbf{z}_s^0(t)=\mathbf{0}$), such that $E[\| \mathbf{z}(t)-\mathbf{z}_s^0(t) \|^4]<C \lambda^{t-s}, \ C<\infty, |\lambda|<1$.
\item[S2:] The data generating system is described by $y(t)=\mathbf{C}\mathbf{x}(t)+ w(t)$, where $\mathbf{x}(t)$ is generated by sampling of the states of a continuously differentiable, bounded and exponentially stable ODE, and where $w(t)$ is generated from a sequence of i.i.d random vectors independent of $\{ \mathbf{u}(t) \} $, by asymptotically stable filtering.
\end{itemize}

\subsection{The Convergence Analysis Tool}
Since there is no projection algorithm defined for ADAM, the boundedness condition needs to be included as an assumption, see \cite{Ljung77a}. The boundedness condition is related to time varying exponential stability, which can be secured by a projection algorithm in combination with a limitation of the adaptation rate, \cite{Wigren94}. The boundedness condition is given by:

{\em The Boundedness Condition}: There is a random variable $\mathcal{C}$ and an {\em infinite} subsequence $\{ t_k \}$, such that $\left( \hat{\bm{\theta}}^{\top}(t_k) \ \ \mathbf{v}^{\top}(t_k) \right)^{\top} \in \bar{\mathcal{D}}_{\mathcal{M}} \subset \mathcal{D}_{\mathcal{M}} \setminus \partial \mathcal{D}_{\mathcal{M}} $ and with $\hat{\mathbf{x}}(t_k)$, $\bm{\Psi}(t_k)$, $\bm{\psi}(t_k)$, $\mathbf{x}(t_k)$, $\mathbf{u}(t_k)$, $w(t_k)$ bounded by $\mathcal{C}$, $\forall t_k$, w.p.1.

Theorem 1 now follows from \cite{Ljung77a} and \cite{Ljung75}:

{\em Theorem 1:} Consider (\ref{eq:eq10}) and assume that M1-M4, G1, A1, A2, S1, S2 and the boundedness condition hold. Also assume that there exists a twice differentiable positive function $V(\bm{\theta},\mathbf{v})$ such that
\begin{eqnarray*}
\frac{d}{d \tau}V(\bm{\theta}_D(\tau), \mathbf{v}_D(\tau)) \leq 0,
\end{eqnarray*}
for $\left(\bm{\theta}_D^{\top}(\tau) \ \ \mathbf{v}_D^{\top}(\tau) \right)^{\top} \in \mathcal{D}_M \setminus \partial \mathcal{D}_M$
when evaluated along solutions of the associated system of ODEs
\begin{eqnarray*}
\frac{d}{d\tau}\bm{\theta}_D(\tau) = - \bar{\alpha}  \mathbf{f}(\bm{\theta}_D(\tau), \mathbf{v}_D(\tau)),
\end{eqnarray*}
\begin{eqnarray*}
\frac{d}{d\tau}\mathbf{v}_D(\tau)= \mathbf{G}(\bm{\theta}_D(\tau)) -\mathbf{v}(\bm{\theta}_D(\tau)).
\end{eqnarray*}
Then
\begin{eqnarray*}
\left(
\hat{\bm{\theta}}^{\top}(t) \ \ \mathbf{v}^{\top}(t) \right)^{\top} \rightarrow \mathcal{D}_C
\end{eqnarray*}
\begin{eqnarray*}
=   \left\{ \left( \bm{\theta}_D^{\top}(\tau) \ \ \mathbf{v}_D^{\top}(\tau) \right)^{\top} \in \mathcal{D}_M \setminus \partial \mathcal{D}_M \right.
\end{eqnarray*}
\begin{eqnarray*}
\left. | \ \frac{d}{d \tau} V(\bm{\theta}_D(\tau),\mathbf{v}_D(\tau))=0 \right\}
\end{eqnarray*}
$ {\rm w.p.1 \ as} \ t \rightarrow \infty$, or
 $\left( \hat{\bm{\theta}}^{\top}(t) \ \ \mathbf{v}^{\top}(t) \right)^{\top} \rightarrow \partial \mathcal{D}_M$.

{\em Proof:} The proof is omitted due to page constraints and since it parallels the corresponding proof of the downloadable open access paper \cite{Wigren23}, with minor changes. The proof of \cite{Wigren23} is pre-ceeded by \cite{Brus08}, supervised by the first author.

\section{Global Convergence}
ADAM is then analysed for two hyper-parameter settings.

\subsection{The Normalized Stochastic Gradient Behaviour}
The first case with close to standard filtering is defined by
\begin{itemize}
    \item[A3:] $\epsilon \rightarrow 0$ and $\beta_2 \rightarrow 1$.
\end{itemize}

First it follows from (\ref{eq:eq200}) and M1 that
\begin{eqnarray*}
\lim_{\beta_2 \rightarrow 1}\lim_{t\rightarrow \infty} \mathbf{v}(t,\bm{\theta})
\end{eqnarray*}
\begin{equation}
\label{eq:eq102}
=\lim_{t \rightarrow \infty}E\left[ \varepsilon^2(t,\bm{\theta}) {\rm \bf{vec}}{\rm (\bf{diag}}(\bm{\psi}(t,\bm{\theta}) \bm{\psi}^{\top}(t,\bm{\theta})))  \right].
\end{equation}

Then consider $\mathbf{f}(\bm{\theta},\mathbf{v})$. An analysis of the element-wise operations of $\left( \mathbf{v}^{\cdot \frac{1}{2}}+\epsilon \mathbf{I} \right)^{\cdot -1}$ shows that the quantity transforms as follows when moved before the expectation
\begin{eqnarray*}
\lim_{\epsilon \rightarrow 0} \lim_{\beta_2 \rightarrow 1}\mathbf{f}(\bm{\theta}, \mathbf{v})
\end{eqnarray*}
\begin{eqnarray*}
= -\left(
\lim_{t \rightarrow \infty}E\left[ \varepsilon^2(t,\bm{\theta}) {\rm (\bf{diag}}(\bm{\psi}(t,\bm{\theta}) \bm{\psi}^{\top}(t,\bm{\theta})))  \right]
\right)^{-\frac{1}{2}}
\end{eqnarray*}
\begin{equation}
\times
\lim_{t \rightarrow \infty} E\left[ \mathbf{m}(t, \bm{\theta}) \right].
\label{eq:eq16}
\end{equation}
This follows since $\lim_{t \rightarrow \infty} \left(1-\beta_1^{{\rm round}(t/T_s)}\right)^{-1}=1$ implies that $\lim_{t \rightarrow \infty}\hat{\mathbf{m}}(t,\bm{\theta})$$=$$\lim_{t \rightarrow \infty}\mathbf{m}(t,\bm{\theta})$. The unity gain of the autoregressive filtering of $\mathbf{m}(t, \bm{\theta})$ in (\ref{eq:eq10}) then gives
\begin{eqnarray*}
\lim_{t \rightarrow \infty}E\left[ \mathbf{m}(t,\bm{\theta})\right]=\beta_1 \lim_{t \rightarrow \infty}E\left[\mathbf{m}(t,\bm{\theta})\right]
\end{eqnarray*}
\begin{equation}
\label{eq:eq203}
+(1-\beta_1)
\lim_{t \rightarrow \infty}E \left[ -\bm{\psi}(t,\bm{\theta}) \varepsilon(t, \bm{\theta})  \right].
\end{equation}
Since $\beta_1$ of (\ref{eq:eq10}) always fulfils $0<\beta_1<1$, it follows that
\begin{equation}
\label{eq:eq15a}
\lim_{t \rightarrow \infty}E \left[ \mathbf{m}(t,\bm{\theta)} \right]=-\lim_{t\rightarrow \infty}E \left[ \bm{\psi}(t,\bm{\theta}) \varepsilon(t,\bm{\theta})  \right].
\end{equation}
When  (\ref{eq:eq15a}) is inserted in (\ref{eq:eq16}) the result is
\begin{eqnarray*}
\lim_{\epsilon \rightarrow 0} \lim_{\beta_2 \rightarrow 1}\mathbf{f}(\bm{\theta}, \mathbf{v})
=\mathbf{f}(\bm{\theta})
\end{eqnarray*}
\begin{eqnarray*}
= -\left(
\lim_{t \rightarrow \infty}E\left[ \varepsilon^2(t,\bm{\theta}) {\rm (\bf{diag}}(\bm{\psi}(t,\bm{\theta}) \bm{\psi}^{\top}(t,\bm{\theta})))  \right]
\right)^{-\frac{1}{2}}
\end{eqnarray*}
\begin{equation}
\times
\lim_{t \rightarrow \infty} E\left[ \bm{\psi}(t, \bm{\theta}) \varepsilon (t, \bm{\theta}) \right],
\label{eq:eq16a}
\end{equation}
where the diagonal matrix is positive definite by M3. A comparison of (\ref{eq:eq16}) with the average updating direction of a steepest descent gradient algorithm, see e.g. \cite{LjungSoderstrom83}, then gives:

{\em Theorem 2:} Assume that M1-M4, A1-A3, S1, S2 and the boundedness condition hold. Then the asymptotic behaviour of the parameter update of (\ref{eq:eq10}) coincides with that of a stochastic gradient algorithm with diagonal normalization.

\subsection{The Asymptotic Sign-Sign Behaviour}
The second case with filtering turned off assumes
\begin{itemize}
    \item[A4:] $\epsilon \rightarrow 0$, $\beta_1= 0$ and $\beta_2= 0$.
\end{itemize}

The turned off filtering does not represent recommended hyper-parameters. However, the analysis contributes to an understanding of the behaviour of ADAM for hyper-parameter settings in between turned off and standard filtering.

In this case $\mathbf{f}(\bm{\theta},\mathbf{v})$ is evaluated by direct simplification, without consideration of (\ref{eq:eq200}). Instead A5 replaces A2:
\begin{itemize}
\item[A5:] The following limit exists for $\bm{\theta} \in \mathcal{D}_\mathcal{M} \subset \mathcal{R}^d$ when $\beta_1= 0$ and $\beta_2 = 0$:
    \begin{eqnarray*}
    \mathbf{f}(\bm{\theta},\mathbf{v})= \lim_{t \rightarrow \infty} E\left[ \left( \mathbf{v}^{\cdot \frac{1}{2}}(t,\bm{\theta}) +\epsilon \mathbf{I} \right)^{\cdot -1} \cdot \hat{\mathbf{m}}(t,\bm{\theta}) \right].
    \end{eqnarray*}
\end{itemize}

Application of  A4 in $\mathbf{f}(\bm{\theta},\mathbf{v})$ of A5, and using M1 gives
\begin{eqnarray*}
 \lim_{\epsilon \rightarrow 0} \mathbf{f}(\bm{\theta},\mathbf{v})=\mathbf{f}(\bm{\theta})=-\lim_{t \rightarrow \infty} E \left[ \right.
\end{eqnarray*}
\begin{eqnarray*}
\left. \frac{ \varepsilon(t,\bm{\theta})}
{\sqrt{\left( \varepsilon(t,\bm{\theta}) \right)^2}}
\left(  {\rm \bf{vec} ( \bf{diag}}(\bm{\psi}(t,\bm{\theta}) \bm{\psi}^{\top}(t,\bm{\theta}))) \right)^{\cdot - \frac{1}{2}} \cdot \bm{\psi}(t,\bm{\theta})  \right]
\end{eqnarray*}
\begin{equation}
=-\lim_{t \rightarrow \infty} E \left[{\rm sign}(\varepsilon(t,\bm{\theta}))  {\rm \bf{sign}}(\cdot \bm{\psi}(t,\bm{\theta}))  \right].
\label{eq:eq11}
\end{equation}
The result (\ref{eq:eq11}) is summarized in:

{\em Theorem 3:} Assume that M1-M4, A1, A4, A5, S1, S2 and the boundedness condition hold. Then the asymptotic behaviour of the parameter update of (\ref{eq:eq10}) coincides with that of a stochastic gradient sign-sign algorithm.

The  sign-sign behaviour is related to the fact that ADAM adapts and normalizes the parameter update for the complete gradient $ \bm{\psi}(t,\bm{\theta}) \varepsilon(t,\bm{\theta})$ in an element-wise way, contrary to Gauss-Newton algorithms, \cite{LjungSoderstrom83}. Referring to \cite{DasguptaJohnsson86} and \cite{Treichleretal87}, it is well known that sign-sign algorithms converge significantly slower than  stochastic gradient algorithms.

\subsection{Global Convergence - Common Part }
The global convergence  analysis of both hyper-parameter cases above is based on the Lyapunov function candidate
\begin{equation}
V(\bm{\theta},\mathbf{v})=V_A(\bm{\theta})= \frac{1}{2}\lim_{t\rightarrow \infty}E[\bm{\varepsilon}^{2}(t,\bm{\theta})]\geq 0.
\label{eq:eq20}
\end{equation}
Since $\mathbf{f}(\bm{\theta},\mathbf{v})$ of (\ref{eq:eq16a}), (\ref{eq:eq11}) and A2 no longer depend on $\mathbf{v}$, and since the associated ODE for $\mathbf{v}$ of A2 is linear and asymptotically stable, it is sufficient to use (\ref{eq:eq20}). Using M1-M4, G1, A1, A2, S1 and S2, the time derivative of the Lyapunov function candidate along the solutions of the associated differential equations of Theorem 1 becomes
\begin{eqnarray*}
\frac{dV(\bm{\theta}_D(\tau),\mathbf{v}_D(\tau))}{d\tau}=\frac{d}{d\tau}
\lim_{t\rightarrow \infty}\frac{1}{2} E \left[ \varepsilon^2(t,\bm{\theta}_D(\tau)) \right]
\end{eqnarray*}
\begin{eqnarray*}
=\lim_{t \rightarrow \infty} \frac{1}{2} E\left[   \frac{\partial \varepsilon^2(t,\bm{\theta})}{\partial \bm{\theta} }
\right]_{|\bm{\theta}=\bm{\theta}_D(\tau)}  \frac{d \bm{\theta}_D(\tau)}{d\tau}
\end{eqnarray*}
\begin{eqnarray*}
=\lim_{t \rightarrow \infty}E \left[
-\bm{\psi}^{\top}(t,\bm{\theta}) \varepsilon(t,\bm{\theta})
\right]_{|\bm{\theta}=\bm{\theta}_D(\tau)} \left(  -\bar{\alpha} \mathbf{f}(\bm{\theta}_D(\tau))\right)
\end{eqnarray*}
\begin{equation}
=\bar{\alpha}\mathbf{f}^{\top}(\bm{\theta}_D(\tau))\lim_{t \rightarrow \infty}E \left[
\bm{\psi}(t,\bm{\theta}) \varepsilon(t,\bm{\theta}) \right]_{|\bm{\theta}=\bm{\theta}_D(\tau)}.
\label{eq:eq17}
\end{equation}

\subsection{Global Convergence in the Normalized Stochastic Gradient Case}
Proceeding from (\ref{eq:eq17}) and using (\ref{eq:eq16a}) immediately gives
\begin{eqnarray*}
\frac{dV(\bm{\theta}_D(\tau),\mathbf{v}_D(\tau))}{d\tau}
\end{eqnarray*}
\begin{eqnarray*}
=-\bar{\alpha} \left( \lim_{t \rightarrow \infty} E\left[ \bm{\psi}(t, \bm{\theta}) \varepsilon (t, \bm{\theta}) \right] \right)^{\top}_{|\bm{\theta}=\bm{\theta}_D(\tau)}
\end{eqnarray*}
\begin{eqnarray*}
\times
\left(
\lim_{t \rightarrow \infty}E\left[ \varepsilon^2(t,\bm{\theta}) {\rm (\bf{diag}}(\bm{\psi}(t,\bm{\theta}) \bm{\psi}^{\top}(t,\bm{\theta})))  \right]
\right)^{-\frac{1}{2}}_{|\bm{\theta}=\bm{\theta}_D(\tau)}
\end{eqnarray*}
\begin{equation}
\times\lim_{t \rightarrow \infty} E\left[ \bm{\psi}(t, \bm{\theta}) \varepsilon (t, \bm{\theta}) \right]_{|\bm{\theta}=\bm{\theta}_D(\tau)} \leq 0.
\label{eq:eq30}
\end{equation}
Equality holds if and only if $\lim_{t \rightarrow \infty} E\left[ \bm{\psi}(t, \bm{\theta}) \varepsilon (t, \bm{\theta}) \right]=0$, referring to M3 and G1.

\subsection{Global Convergence and the Symmetry Requirement in the Sign-Sign Case}
Combining (\ref{eq:eq17}) and (\ref{eq:eq11}) gives
\begin{eqnarray*}
\frac{dV(\bm{\theta}_D(\tau),\mathbf{v}_D(\tau))}{d\tau}
\end{eqnarray*}
\begin{eqnarray*}
=-\bar{\alpha}\lim_{t\rightarrow \infty}E \left[
{\rm sign}(\varepsilon(t,\bm{\theta})){\rm \bf{sign}}\left(\cdot \bm{\psi}^{\top}(t,\bm{\theta}) \right)
\right]_{|\bm{\theta}=\bm{\theta}_D(\tau)}
\end{eqnarray*}
\begin{eqnarray*}
\times
\lim_{t \rightarrow \infty} E \left[ \bm{\psi}(t, \bm{\theta}) \varepsilon(t,\bm{\theta})
\right]_{|\bm{\theta}=\bm{\theta}_D(\tau)}
\end{eqnarray*}
\begin{eqnarray*}
=-\bar{\alpha}\lim_{t\rightarrow \infty}E \left[
{\rm \bf{sign}}\left( \cdot \bm{\psi}(t,\bm{\theta})\varepsilon(t,\bm{\theta})) \right)^{\top}
\right]_{|\bm{\theta}=\bm{\theta}_D(\tau)}
\end{eqnarray*}
\begin{equation}
\times
\lim_{t \rightarrow \infty} E \left[ \bm{\psi}(t, \bm{\theta}) \varepsilon(t,\bm{\theta})
\right]_{|\bm{\theta}=\bm{\theta}_D(\tau)}.
\label{eq:eq21}
\end{equation}
The following assumption is now introduced
\begin{itemize}
\item[A6:] The distribution of the components of the stochastic vector variable $\bm{\psi}(t,\bm{\theta}) \varepsilon(t,\bm{\theta})$ are symmetric around their mean values when $t\rightarrow \infty$.
\end{itemize}
The reason why A6 is introduced is the following result:

{\em Lemma 1:} Assume that the distribution $p_X$ of the stochastic variable $X$ is symmetric around its mean $\bar{x}$. Then $E\left[ {\rm sign}(X) \right]= {\rm sign} \left( E \left[ X \right]  \right)$.

{\em Proof:} The proof is by direct calculation.
\begin{eqnarray*}
E \left[ {\rm sign}(X) \right]=\int_{-\infty}^{\infty} {\rm sign}(x)p_X(x)dx
\end{eqnarray*}
\begin{eqnarray*}
=\int_{-\infty}^{0} (-1)p_X(x)dx+\int_{0}^{\infty} (1)p_X(x)dx
\end{eqnarray*}
\begin{eqnarray*}
=\int_{-\bar{x}}^{\infty}  p_X(z+\bar{x})dz - \int_{-\infty}^{-\bar{x}} p_X(z+\bar{x})dz
\end{eqnarray*}

\begin{eqnarray*}
=\int_{-\bar{x}}^{\bar{x}} p_X(z+\bar{x})dz.
\end{eqnarray*}
Noting that the integral is positive if $\bar{x}>0$ and negative if $\bar{x}<0$, Lemma 1 follows.

A6 and an element-wise use of Lemma 1 in
 (\ref{eq:eq21}) gives
\begin{eqnarray*}
\frac{dV(\bm{\theta}_D(\tau),\mathbf{v}_D(\tau))}{d\tau}
\end{eqnarray*}
\begin{eqnarray*}
=-\bar{\alpha}
{\rm \bf{sign}} \left( \cdot \lim_{t\rightarrow \infty}E \left[
\left( \bm{\psi}(t,\bm{\theta})\varepsilon(t,\bm{\theta}) \right)^{\top}
\right]_{|\bm{\theta}=\bm{\theta}_D(\tau)}  \right)
\end{eqnarray*}
\begin{eqnarray*}
\times
  \lim_{t \rightarrow \infty} E \left[ \bm{\psi}(t, \bm{\theta}) \varepsilon(t,\bm{\theta})
\right]_{|\bm{\theta}=\bm{\theta}_D(\tau)}
\end{eqnarray*}
\begin{equation}
=-\bar{\alpha} \sum_{i}| \lim_{t \rightarrow \infty} E \left[ \psi_i(t, \bm{\theta}) \varepsilon(t,\bm{\theta})
\right]_{|\bm{\theta}=\bm{\theta}_D(\tau)} | \leq 0.
\label{eq:eq23}
\end{equation}
Again, equality holds if and only if $\lim_{t \rightarrow \infty} E \left[ \bm{\psi}(t, \bm{\theta}) \varepsilon(t,\bm{\theta})
\right]_{|\bm{\theta}=\bm{\theta}_D(\tau)}=0$.

\subsection{Global Convergence to the True System}

It is now noted that the derived condition for global convergence for both hyper-parameter settings is given by $\lim_{t \rightarrow \infty} E \left[ \bm{\psi}(t, \bm{\theta}) \varepsilon(t,\bm{\theta})
\right]_{|\bm{\theta}=\bm{\theta}_D(\tau)}=0$.  The following assumption is therefore convenient
\begin{itemize}
\item[S3:] There exist  parameter vectors $\bm{\theta}^{\star}$ such that $y(t)=\hat{y}(t,\bm{\theta}^{\star})+\varepsilon(t,\bm{\theta}^{\star})$, where $\varepsilon(t,\bm{\theta}^{\star})$ is independent of $\mathbf{u}(t)$, with zero mean.
\end{itemize}
By S3, the condition $\lim_{t \rightarrow \infty} E \left[ \bm{\psi}(t, \bm{\theta}) \varepsilon(t,\bm{\theta})
\right]_{|\bm{\theta}=\bm{\theta}_D(\tau)}=0$ holds for all $\bm{\theta}^{\star}$ since $\bm{\psi}(t, \bm{\theta})$ is generated only from $\mathbf{u}(t)$ which is independent of $\varepsilon(t,\bm{\theta})$.
Theorem 1 now implies:

{\em Theorem 4:} Assume that the boundedness condition, M1-M4, G1, A1, and S1-S3 hold for (\ref{eq:eq10}). If i) A2 and A3 hold, or ii) A4, A5 and A6 hold, then
$ \hat{\bm{\theta}}(t) \rightarrow \mathcal{D}_C \ {\rm w.p.1} \ {\rm as} \ t \rightarrow \infty$,
or $\hat{\bm{\theta}}(t)  \rightarrow \partial \mathcal{D}_M$, where
$\bm{\theta}^{\star} \in \mathcal{D}_C$ is defined in Theorem 1.

Convergence is global and Theorem 4 is valid for both cases treated by the paper. However there may be other sub-optimal classes of points in the invariant set $\mathcal{D}_C$ than $\bm{\theta}^{\star}$, cf. e.g. \cite{Reddietal18}. Such sub-optimal points do then not meet S3, but they do fulfil $\lim_{t \rightarrow \infty} E \left[ \bm{\psi}(t, \bm{\theta}) \varepsilon(t,\bm{\theta})
\right]_{|\bm{\theta}=\bm{\theta}_D(\tau)}=0$. Note also that S3 implies that $w(t)$ of S2 can replace $\varepsilon(t,\bm{\theta}^{\star})$.

\section{Numerical Results}

To test the proposed RNN and to validate the results of the averaging analysis, a Python-based Monte-Carlo analysis of a simulated automotive cruise control system was performed. The vehicle traveling with velocity $x_1(t)$ is subject to  thrust, friction, air resistance and gravitational forces in hilly terrain, see  e.g. \cite{Nilssonetal16}. Here, the friction and gravitational forces are treated as a disturbance $w(t)$. Newton's second law gives
\begin{equation}
\dot{x}_1(t)=u(t)-\frac{\rho A C_{x_1}}{2m}x_1^2(t)-w(t),
\label{eq:eq100}
\end{equation}
In (\ref{eq:eq100}), $u(t)$ is the accelerator command, $m$  the mass of the vehicle, $A$ the frontal area, $\rho$ the density of the air, and $C_{x_1}$ is the air resistance coefficient. This system was sampled with  $T_s=0.1$ $s$. The mass of the vehicle was $m=1500$ $kg$, while $\rho$, $A$ and $C_{x_1}$ were set to give the vehicle a maximum speed of $60$ $m/s$. The maximum acceleration and retardation were $\pm 3.0$ $m/s^2$. The white velocity measurement standard deviation was $0.1$ $m/s$, while the standard deviation of $w(t)$ was $0.01$ $m/s^2$.  To identify the dynamics, $f^{nn}\left(\hat{\mathbf{x}}(t,\bm{\theta}),\mathbf{u}(t), \bm{\theta} \right)$ was selected with one hidden layer of width 8. The analysis averaged twenty runs for each of the two hyper-parameter settings analysed in Section IV, using fix $\alpha$ values $0.001$ and $0.0001$ to also illustrate tracking. The $tanh $ activation functions was used. The results in Fig. 1 and Fig. 2 are consistent with the analysis, with the standard hyper-parameter setting performing significantly better than the sign-sign one. The sign-sign hyper-parameter setting with $\alpha=0.0001$ leads to very slow convergence. This may be due to the algorithm, a suboptimal $\bm{\theta}^{\star}$, or that A5 fails to hold.
\begin{figure}[t]
\begin{centering}
\includegraphics[width=2.9in]{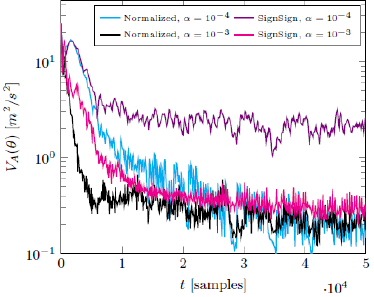}
\caption{Simulated convergence speeds of ADAM.}
\end{centering}
\label{fig:fig1}
\end{figure}

\begin{figure}[t]
\begin{centering}
\includegraphics[width=2.7in]{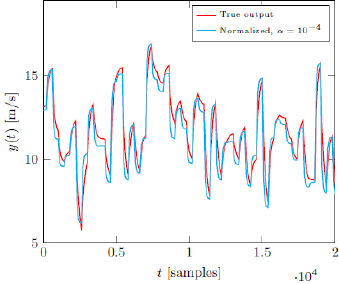}
\caption{True and predicted output after training.}
\end{centering}
\label{fig:fig2}
\end{figure}

\section{Conclusions}
The paper derived the asymptotic average updating direction of ADAM for two  hyper-parameter settings. It was proved that the setting that represents close to standard hyper-parameters behaves as a diagonally power normalized stochastic gradient algorithm. The case with filtering turned off instead behaves as a stochastic sign-sign algorithm. In addition it was proved that the algorithm converges globally to an invariant set that is a superset of the parameter vectors that represent perfect input-output models. The paper also proposed a model structure that embeds structure in RNNs. A Monte-Carlo simulation study validated the results. In view of the asymptotic similarity to other diagonally power scaled gradient descent algorithms, \cite{Makinoetal93}, \cite{Wigren98}, a significant performance advantage for ADAM with respect to conventional normalized gradient descent algorithms is expected, cf. \cite{Wigren98}.

\end{document}